# Cross-Dataset Generalization of Glaucoma Detection CNNs: Effects of Operating Threshold, Multi-Source Training, and Color Normalization

Al-Amin Rahman*
Department of Electrical and Electronic Engineering
alaminrahman.eee.33@gmail.com

Saifuddin Ahmed
Department of Electrical and Electronic Engineering
sec.saifuddin@gmail.com

Resalat Ahmed
Department of Electrical and Electronic Engineering
ahmedresalat35@gmail.com

Sylhet Engineering College, Sylhet, Bangladesh
(Affiliated with Shahjalal University of Science & Technology, Sylhet)

***Abstract*—Deep learning models for glaucoma detection are often tested on images from the same dataset used for training. This does not show how the model will work at a new hospital or with a new camera. We studied this problem using five ImageNet-pre-trained CNNs (ResNet50, DenseNet121, MobileNetV2, EfficientNet-B0, and InceptionV3). Each was fine-tuned with identical settings on one of three public fundus datasets (ACRIMA, ORIGA, and RIM-ONE DL), then tested on a held-out part of its own dataset and on the other two datasets in full. All experiments used three random seeds, and we report mean ± SD over seeds. We also examined threshold recalibration, multi-source training, four image normalization methods, and stronger color augmentation to determine whether these approaches could improve transfer performance. The mean in-domain AUC was 0.913, compared with 0.702 for cross-source testing, a drop of 0.211 (SD 0.030; 95% CI 0.135 to 0.286). After correction for multiple comparisons, we found no evidence that any architecture was better than the others, although three seeds give low power. At thresholds chosen on the source validation sets, mean specificity fell from 0.870 to 0.506 and mean sensitivity from 0.831 to 0.726. Recalibrating the threshold on 10% of the target images raised specificity to 0.670 but lowered sensitivity to 0.609, and balanced accuracy rose only from 0.616 to 0.639. Multi-source training did not improve mean held-out AUC compared with single-source training (0.695 versus 0.702). Among the four normalization methods and a stronger color augmentation, LAB reference normalization (cross-source AUC 0.755; paired gain 0.053, 95% CI 0.046 to 0.060 across seeds) and strong color augmentation (0.754; wider brightness, contrast, saturation and hue changes) gave similar results, reducing the drop to 0.155 and 0.158. Ben Graham normalization (0.733) and CLAHE (0.712) gave smaller gains, and gray-world normalization (0.696) gave none. The two best methods raised specificity but lowered sensitivity. On a fourth dataset, G1020, used only for testing, AUC stayed between 0.492 and 0.502 in every condition, and networks trained and tested within G1020 were also close to chance. These findings show that strong in-domain performance does not guarantee reliable cross-dataset glaucoma detection, and that normalization and color augmentation reduce the drop but do not remove it. Studies should include external validation and a check of the operating threshold, not only discrimination.**

## I. Introduction

Glaucoma is a disease of the optic nerve. It is one of the main causes of permanent blindness in the world [1], [2]. Tham *et al.* estimated that 64.3 million people aged 40 to 80 years had glaucoma in 2013. They expected this number to reach 111.8 million by 2040 [1]. The damage cannot be reversed. In the early stage there are usually no symptoms, so the disease is found by looking at the optic nerve head [2]. A color fundus photograph shows the optic disc at low cost. Doctors look at the size of the cup compared with the disc (the cup-to-disc ratio), the shape of the rim and the nerve fiber layer. Grading photographs by hand takes time and depends on the skill of the grader. For this reason, automatic screening is attractive, especially where there are few specialists.

Deep learning has shown good agreement with doctors in diabetic retinopathy screening [3]. Many groups have also used convolutional neural networks (CNNs) for glaucoma [4]–[7]. Reported AUC values are often above 0.95. In many of these papers, the training and test images come from the same dataset. They share the same camera, the same hospital, the same cropping and the same labeling habits. A screening tool will not be used under these conditions. It will get images from other devices and other patients. A model that was only checked on its own dataset may behave very differently there. Drops of this kind were found for chest radiographs [8]. The problem of dataset bias in image classification is older than deep learning [9].

In this study we ask a simple question. If a standard ImageNet pre-trained CNN is fine-tuned on one public glaucoma dataset, how much AUC is lost on images from another public dataset? Do some architectures lose less than others? We also ask whether image normalization or a stronger color augmentation can reduce the loss. We do not propose a new network. We use five well-known architectures from one library, train them with one fixed protocol, and report what happens when the test images come from a different source.

The contributions of this paper are as follows.

1) A comparison of five pre-trained CNNs under identical training settings (same optimizer, learning rate, epochs and $224 \times 224$ input, with no per-model tuning). Each model was trained on each of three public datasets and tested in-domain and on the other datasets, with three random seeds. Results are given as mean ± SD over seeds.

*Corresponding author: alaminrahman.eee.33@gmail.com

2) A measurement of the AUC drop from in-domain to cross-source testing for each model and each source dataset. We also report sensitivity and specificity at the threshold chosen on the source validation set.
3) A check of the G1020 dataset. We computed the cup-to-disc ratio from its masks and also trained and tested the networks inside G1020. G1020 was close to unpredictable for all models, so we use it only as a separate test set.
4) A test of threshold recalibration. The threshold of each trained network was set again on a small labeled sample (10%) of the target dataset. We compared sensitivity and specificity with the source threshold on the remaining target images.
5) A leave-one-source-out test of multi-source training. A network was trained on two datasets together and tested on the third. A size-matched control separates the effect of more data from the effect of more varied data.
6) A comparison of gray-world, Ben Graham, CLAHE and LAB reference normalization with a stronger color augmentation. We report paired seed-level AUC changes, sensitivity and specificity, source-target effects and a separate G1020 test.

The paper is organized as follows. Section II gives related work. Section III describes the datasets, models and protocol. Section IV gives the results. Section V discusses them and the limits of the study. Section VI concludes.

## II. Related Work

### A. Deep Learning for Glaucoma from Fundus Images

Chen *et al.* [4] were among the first to classify glaucoma from fundus images with a CNN. They tested it on the ORIGA and SCES collections. Later work used deeper networks and larger image sets. Li *et al.* [5] trained a deep learning system on a large number of graded color fundus photographs. Gómez-Valverde *et al.* [6] used transfer learning with a pre-trained CNN on images from several sources. Diaz-Pinto *et al.* [7] compared ImageNet pre-trained networks on public databases. They also released the ACRIMA database, which we use. Fu *et al.* [10] proposed a disc-aware ensemble network that uses both the whole image and the disc region. The REFUGE challenge [11] gave a common dataset with images from more than one device and a common way to evaluate glaucoma classification and disc and cup segmentation.

Some studies did look beyond a single dataset. Diaz-Pinto et al. [7] combined cross-validation with cross-testing, in which models fine-tuned on some public databases were tested on others. Other work treats the shift between glaucoma datasets as a domain adaptation problem. Bao et al. [43] proposed a self-adaptive transfer learning method for multicenter glaucoma classification and evaluated it between a private dataset and two public ones. These studies show that dataset shift matters for glaucoma. Many other studies still test the model on a held-out part of the data used for training, or on a mix of several sources. This shows that CNNs can learn the task. It does not show whether a trained network still works when the images come from a different source. This is the question we study here. We compare several architectures under one fixed protocol with repeated seeds, and we examine the operating threshold as well as the AUC. In addition, we test simple multi-source training and compare image normalization with color augmentation under the same cross-source protocol.

### B. Dataset Shift and Cross-Dataset Evaluation

Torralba and Efros [9] showed that classifiers trained on one popular image dataset lose accuracy on another, even when both have the same object classes. Domain adaptation theory explains part of this. The error on a target domain is limited by the source error plus a term that grows with the difference between the two input distributions [12]. In medical imaging, Zech *et al.* [8] found that a pneumonia model worked worse on radiographs from hospitals that were not used for training. Other studies show that networks can learn shortcuts, such as hospital-specific markers, instead of signs of disease [13], [14]. Such shortcuts do not transfer to a new site. In retinal imaging, Wang *et al.* [15] used adversarial domain adaptation for the shift between fundus datasets in optic disc and cup segmentation.

Transfer learning from ImageNet [16] is the usual starting point for small medical datasets. Yosinski *et al.* [17] studied how general the features of each layer are. Raghu *et al.* [18] found that ImageNet pre-training helps medical images less than expected in some settings, and that small networks can be as good as large ones. For this reason we compare architectures of different sizes under the same conditions, and we test them on other datasets, not only on their own test split.

### C. Color and Illumination Normalization

Image appearance can be changed before classification without changing the network. Gray-world balancing rescales the color channels, assuming that their averages should be equal [19]. Graham used local-average subtraction and intensity rescaling to reduce lighting differences in retinal photographs [20]. Contrast-limited adaptive histogram equalization (CLAHE) improves local contrast and limits the amplification of the histogram [21]. Statistical color transfer moves the color statistics of an image toward those of a reference image [22]. These methods act on different parts of the appearance, so they need not have the same effect on classification.

Color augmentation is a different idea. Random changes of color in training broaden the range of appearances that the network sees. They do not map every image to one fixed appearance. Torchvision offers such changes through ColorJitter [23]. We compare these approaches with the same cross-dataset evaluation. An image that looks more uniform does not by itself prove better glaucoma detection. What counts is the performance on unseen datasets and at the chosen decision threshold.

## III. Methodology

### A. Datasets

We used four public fundus datasets (Table I). ACRIMA [7] has 705 images (396 glaucoma, 309 normal). They were taken at FISABIO Oftalmólogia Médica in Valencia, Spain, and labeled by glaucoma experts. The images are cropped around the optic disc. RIM-ONE DL [24] has 485 images (172 glaucoma, 313 normal) from three Spanish hospitals. ORIGA-light [25] has 650 images (168 glaucoma, 482 normal). G1020 [26] has 1020 images (296 glaucoma, 724 normal). For ORIGA and G1020 we cut a square window around the optic disc. The window was centered on the disc mask supplied with the dataset, and its side was 2.5 times the disc size. Where there was no usable mask, we used automatic disc detection. The crops were saved at 512 × 512 pixels. We used the binary labels (glaucoma or normal) as given by each dataset.

TABLE I
DATASETS USED IN THIS STUDY

| Dataset | Images | Glaucoma / normal | Image content | Role |
|---|---|---|---|---|
| ACRIMA | 705 | 396 / 309 | Disc crop, variable size | Source, target |
| ORIGA | 650 | 168 / 482 | Disc crop (by us), 512×512 | Source, target |
| RIM-ONE DL | 485 | 172 / 313 | Disc-centered, variable size | Source, target |
| G1020 | 1020 | 296 / 724 | Disc crop (by us), 512×512 | Target only |

Fig. 1 shows four random images from each dataset. The datasets differ in several ways. ACRIMA and RIM-ONE DL images are close views of the disc, but their size changes from image to image (ACRIMA samples from 202 to 637 pixels, RIM-ONE DL samples from 401 to 589). The color also differs. ACRIMA images are strongly saturated orange-red. RIM-ONE DL images are darker and redder, with a vignette at the border. ORIGA images are more yellow-brown. The G1020 images show a wider part of the retina, so the optic disc is a smaller part of the image, and some images have black areas from the field of view of the camera. The baseline used no color normalization and no further disc localization, so the networks saw all these differences. Section III-J describes a separate test of whether changing the image appearance reduces their effect.

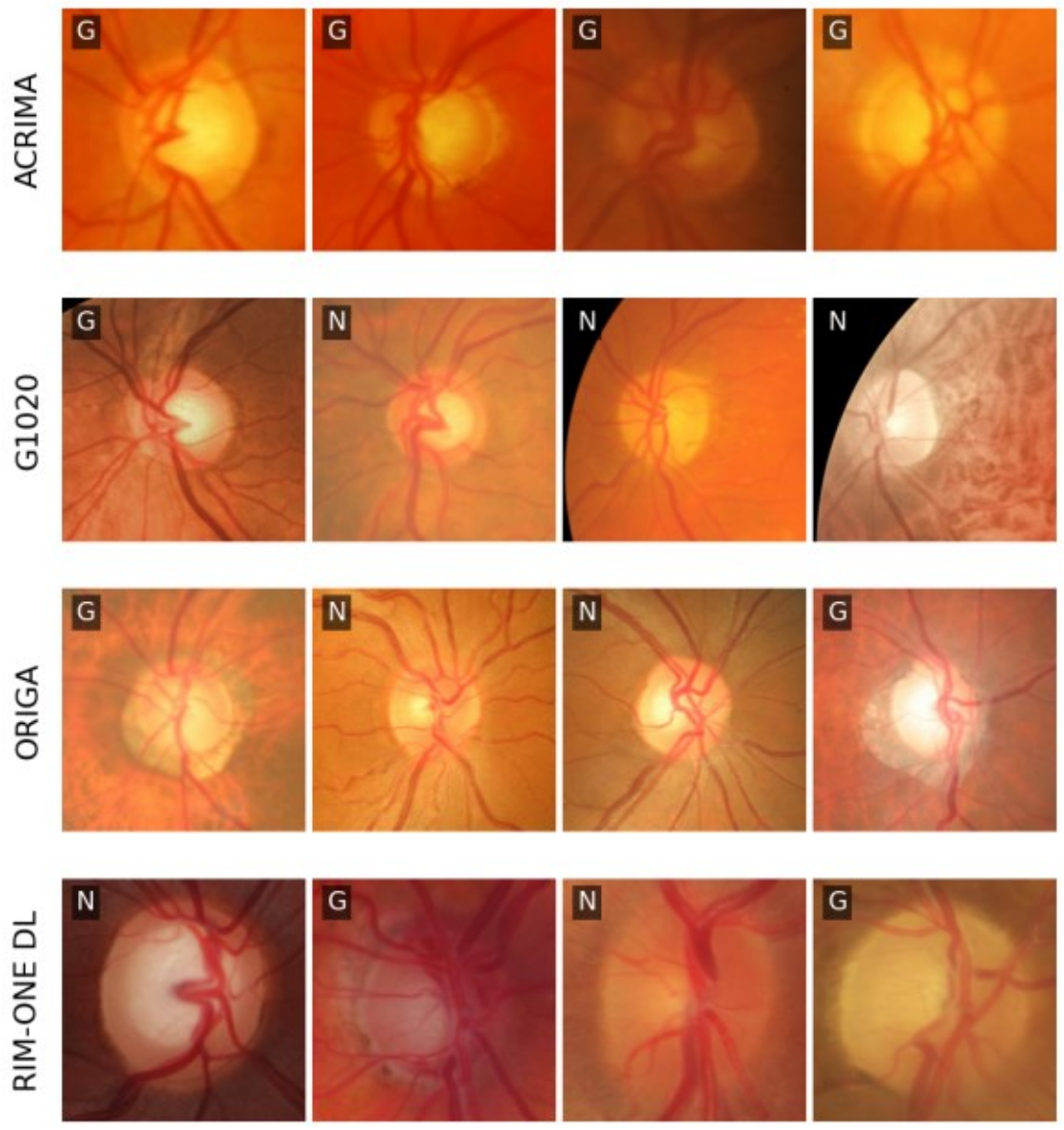

Fig. 1. Four random images from each dataset (rows: ACRIMA, G1020, ORIGA, RIM-ONE DL). G: glaucoma label, N: normal label.

## B. Preprocessing and Augmentation

For the baseline, all images were converted to RGB and resized to 224 × 224 pixels without keeping the aspect ratio. They were then normalized with the ImageNet mean and standard deviation. Training images were augmented with random horizontal flips, random rotations up to 15 degrees, and random brightness and contrast changes up to 20%. Validation and test images were not augmented. The ImageNet standardization only rescales the network input. It is not the same as the image-level color and illumination changes in Section III-J.

## C. Models and Training

We used five CNNs from the torchvision library of PyTorch [27] with ImageNet weights: ResNet50 [28], DenseNet121 [29], MobileNetV2 [30], EfficientNet-B0 [31] and InceptionV3 [32]. In each network the last layer was replaced by one output unit that gives the glaucoma score. The auxiliary classifier of InceptionV3 was removed. All layers were fine-tuned. The settings were the same for every model: AdamW optimizer [33] with learning rate $10^{-4}$ and weight decay $10^{-4}$, cosine annealing of the learning rate over 15 epochs [34], batch size 32, at most 15 epochs, and mixed-precision arithmetic [35]. The loss was binary cross-entropy. The positive class was weighted by the ratio of normal to glaucoma images in the training split, because the classes are not balanced. After each epoch we computed the AUC on the validation split and kept the weights of the best epoch. Training stopped after five epochs without improvement. The decision threshold was chosen on the validation split by maximizing Youden's J statistic [36]. When the validation AUC is 1.0, many thresholds give the same J, so the chosen value is not well determined (Section IV-C). Test images were never used for any of these choices. Training was done on a free Google Colab T4 GPU.

### D. Experimental Protocol

Each of the three primary datasets (ACRIMA, ORIGA, RIM-ONE DL) was used in turn as the source. The source dataset was split into 70% training, 15% validation and 15% test images, stratified by class. The 15% test part gives the in-domain result (106, 98 and 73 images for ACRIMA, ORIGA and RIM-ONE DL). For the cross-source result, the model was tested on the complete other two datasets. None of their images was used for training, validation or threshold selection. Each model was also tested on the complete G1020 set. G1020 was never a source in the main experiments. The only time we trained on it was the separate in-domain check in Section III-F.

The baseline gives 5 models × 3 sources = 15 trained networks per seed. The experiment was repeated with seeds 42, 7 and 123, so 45 networks were trained, and each was tested on four targets. The seed controls the data split, the start values of the new output layer, the batch order and the augmentation. The spread between seeds therefore includes both split and training randomness. The same protocol was used for every run. The normalization and color augmentation comparison is described separately in Section III-J.

### E. Metrics and Aggregation

The main metric is the area under the ROC curve (AUC) [37]. It does not depend on a threshold, so it can be compared between datasets with different class ratios. We also report sensitivity, specificity and accuracy at the stored validation threshold, computed with scikit-learn [38]. For one model and one seed, the in-domain AUC is the mean of the three diagonal cells of the 3 × 3 source-target matrix. The cross-source AUC is the mean of the six off-diagonal cells. The AUC drop is

$$\Delta AUC = AUC_{in} - AUC_{cross} \tag{1}$$

It is computed for each seed first. Only then do we take the mean and the sample standard deviation ($n - 1$) over the three seeds. In this way the seed is the unit of replication, and we never average over images. Sensitivity, specificity and accuracy are summarized in the same way. Confidence intervals of seed-level means use the t distribution with two degrees of freedom. Models are compared with paired t-tests on the seed-level values, with a Bonferroni threshold of 0.005 for the ten model pairs. With three seeds these tests have low power, so they are exploratory. For G1020, the external drop is the in-domain AUC of the source dataset minus the G1020 AUC of the same model and seed. G1020 results are kept apart from the 3 × 3 results, and no four-dataset average is computed.

### F. Checking the G1020 Labels

Before using G1020 as a source, we checked how well its labels agree with the geometry of the optic disc. From the disc and cup masks we computed the vertical cup-to-disc ratio (vCDR, the vertical size of the cup mask divided by that of the disc mask), the horizontal ratio (hCDR) and the area ratio. For 230 of the 1020 images (170 normal, 60 glaucoma), the disc or cup mask was empty, so no ratio could be computed. The analysis uses the other 790 images (554 normal, 236 glaucoma). Each ratio was used as a one-feature classifier and scored by AUC.

As a second check, we trained the same five networks on G1020 itself with the settings of Section III-C. We used a stratified 70/15/15 split (153 test images, 44 with glaucoma) and seeds 42, 7 and 123. These in-domain results are reported separately. They are not part of the 3 × 3 matrix, the external test or any average.

### G. Grad-CAM

To see which image region drives the prediction, we applied Grad-CAM [39] with the pytorch-grad-cam package [40]. We used a ResNet50 fine-tuned on ACRIMA, the last residual block, and eight images from the ACRIMA test split. This is a qualitative check only.

### H. Threshold Recalibration

We tested whether a small labeled sample from the new dataset can repair the threshold. For each model, seed, source and target, 10% of the target images formed the calibration set. They were drawn at random and stratified by class (48 to 70 images for ACRIMA, ORIGA and RIM-ONE DL, and 102 for G1020). The new threshold maximized Youden's J on these images, which is the same rule as for the source threshold. The network weights were not changed. We then computed sensitivity, specificity, accuracy and balanced accuracy (the mean of sensitivity and specificity) on the remaining 90% of the target images. This was done once with the source threshold and once with the new threshold, so both are compared on the same images.

The random split was repeated 200 times, and the metrics were averaged over the repeats within each seed. The repeats are not independent samples, so the SD and the confidence intervals use only the three seeds, as in Section III-E. We repeated the procedure with 5%, 20% and 30% of the target images (100 splits each). All values in this subsection and in Section IV-F come from this run. The AUC does not depend on the threshold, so it is not recalibrated. G1020 was treated in the same way and is reported separately.

### I. Multi-Source Training

We tested whether pooling several sources reduces the loss on an unseen dataset. For each target among ACRIMA, ORIGA and RIM-ONE DL, a network was trained on the pooled training parts of the other two datasets. The pooled validation parts were used to choose the epoch and the Youden threshold. The splits, seeds, models and hyper-parameters were the same as in Sections III-C and III-D. The held-out dataset was never used for training, validation or threshold selection. It was tested in full, so its images are the same as in the single-source cross-source cells. G1020 was tested separately and never entered a pool.

A control arm separates the effect of data volume from the effect of source diversity. In this arm we took a random stratified subsample of each source, so that the pool had as many images as the smaller single-source training set (about 340 to 450 images). For every model, seed and target, the multi-source AUC was compared with the mean of the two single-source AUCs on the same target. For reference we also compared it with the better and the worse of the two. Differences were com-puted for each seed, averaged over the five models and summarized as mean ± SD over seeds 42, 7 and 123, with t-based 95% confidence intervals. Before looking at the results we fixed a rule. An improvement counts as supported only if the confidence interval of the all-model gain excludes zero and the gain is at least 0.02 AUC. The single-source results are those of the original experiment, which was not run again. The multi-source networks were trained once per arm and seed. We also applied the recalibration of Section III-H to the multi-source scores.

### *J. Normalization and Color Augmentation*

We compared the baseline with five changes of image appearance: gray-world channel balancing, Ben Graham local background subtraction, CLAHE, LAB reference normalization, and stronger color augmentation. The first four are fixed preprocessing steps. The last one is augmentation, so we keep it apart from normalization throughout. The list of methods was fixed before the runs, and all of them are reported. No method was chosen by its test AUC. Fig. 2 shows the baseline and the four preprocessing methods on one image from each dataset. We looked at these images before training to check that the optic disc stayed visible. No parameter was changed afterwards.

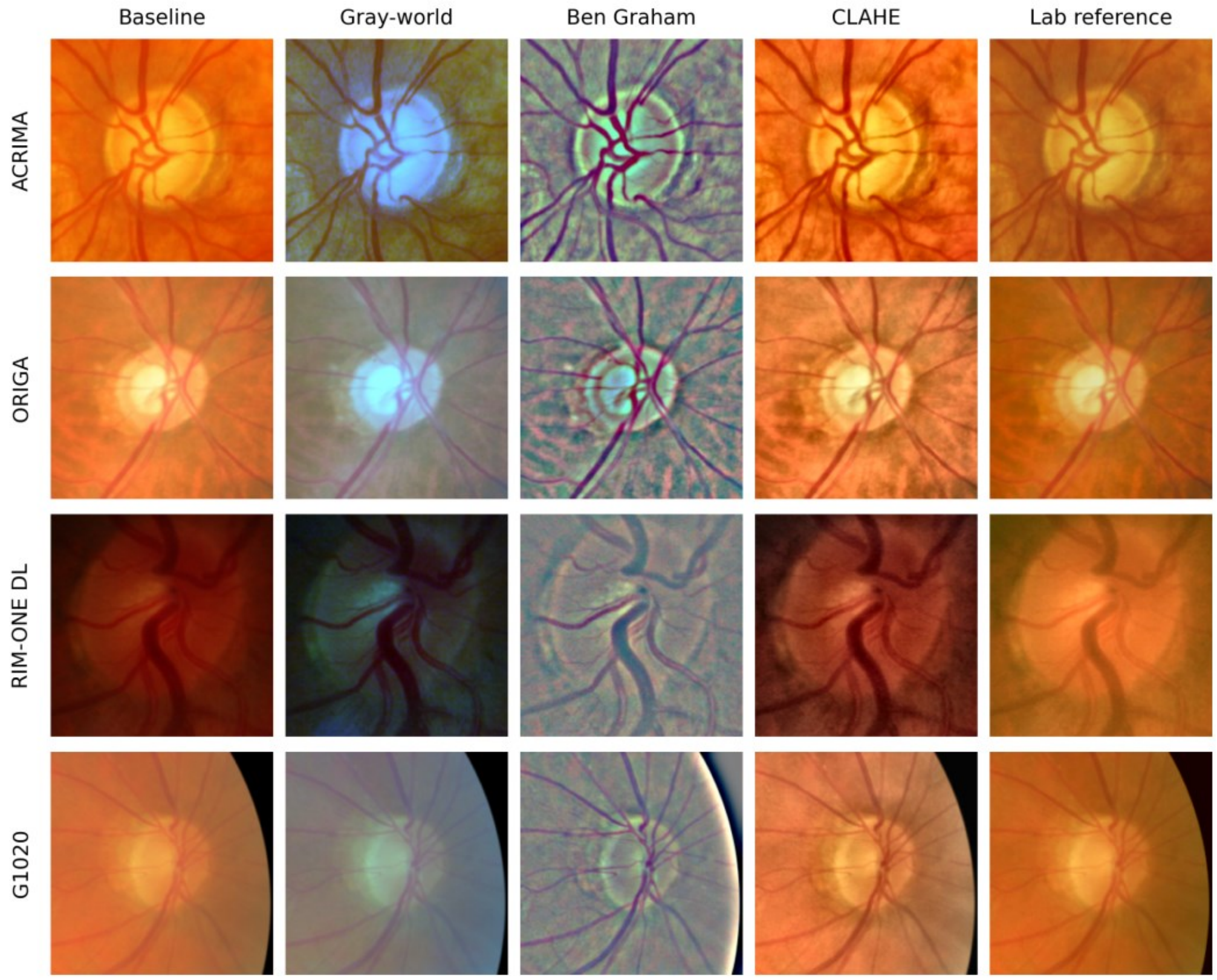

Fig. 2. Image appearance under the baseline and the four preprocessing methods. Rows show one image each from ACRIMA, ORIGA, RIM-ONE DL and G1020. The same image is used across each row. Strong color augmentation is random and is not shown. The LAB reference in this figure was computed from the ACRIMA training images (seed 42).

All preprocessing acts after resizing to 224 × 224 pixels (bilinear interpolation) and before the random augmentation and the ImageNet standardization. In training, the order is resize, normalization, horizontal flip, rotation, color jitter, conversion to a tensor and ImageNet standardization. In validation and testing, it is resize, normalization, conversion to a tensor and ImageNet standardization. The preprocessing uses no label. It is applied to training, validation, test and cross-source images in the same way. The result is clipped to the range 0–255 and stored as 8-bit integers.

We call a pixel a retina pixel if the largest of its R, G and B values is above 15 (scale 0–255). The statistics for gray-world and for the LAB reference are computed on retina pixels only. If 10% or fewer of the pixels qualify, all pixels are used. The correction itself is applied to every pixel. Gray-world multiplies each channel by the mean of the three channels means divided by the mean of that channel [19]. The Ben Graham transform is $4I - 4(G_\sigma * I) + 128$ for each RGB channel [20]. The Gaussian has $\sigma = 224/30$, about 7.5 pixels (OpenCV [41], default border handling, no circular mask). CLAHE [21] is applied to the L channel of the 8-bit CIELAB image with clip limit 2.0 and 8 × 8 tiles. The a and b channels are not changed.

The LAB reference maps the mean and standard deviation of each CIELAB channel of an image to fixed reference values. This is in the spirit of statistical color transfer [22], but it uses CIELAB instead of lαβ and fixed values instead of a reference image. The reference values are the averages of the per-image CIELAB means and standard deviations over the training images of the source dataset only. They are computed again for every source and seed, and they were not tuned. Validation, test and target images and all labels play no part in them. An image from another dataset is mapped with the reference of the source dataset, but only its own statistics are used.

The control replaces the baseline color jitter (brightness and contrast 0.2) by brightness, contrast and saturation factors drawn between 0.6 and 1.4 and a hue shift up to ±0.1. It is used in training only. Flip and rotation are unchanged. It adds saturation and hue changes as well as wider ranges, so it is a package of changes and not only a stronger version of the baseline jitter. In the four preprocessing conditions the baseline jitter is kept. We did not test methods that use unlabeled target images, such as histogram matching to the target or adaptive batch normalization, because they are domain adaptation.

All five networks were trained under each of the five conditions on each source with seeds 42, 7 and 123. This gives 225 additional networks. The splits, seeds, hyper-parameters, threshold selection and evaluation are those of Sections III-C and III-D. G1020 was test-only. The baseline is the original result set, which was not run again. For a given seed, the baseline run and the changed run share the split, the start values of the output layer and the test images. The difference between them therefore shows the effect of the change.

For every model and seed we computed the in-domain AUC, the cross-source AUC and the AUC drop as in Section III-E. The differences from the baseline are paired by model and seed. The all-model result is first averaged over the five architectures within each seed and then summarized over the three seeds. G1020 is summarized separately. Sensitivity, specificity and accuracy are taken at the source-selected thresholds, without recalibration on the target. This is a separate single-source test. It does not combine normalization with pooling or recalibration.

For each method we report the change in in-domain AUC, cross-source AUC and AUC drop. A positive change in cross-source AUC means better transfer. A negative change in the drop means a smaller drop. A smaller drop must be read together with both AUC values, because it can also come from a lower in-domain AUC. Paired differences use t-based 95% confidence intervals with two degrees of freedom. The intervals and p-values are not adjusted. As an exploratory check, we compared the five cross-source contrasts with 0.01 (0.05/5). This does not cover all outcomes or all architecture-specific comparisons together. Cell-level and architecture-level patterns are described, not tested.

# IV. Results

## *A. In-Domain versus Cross-Source Performance*

Fig. 3 shows the 3 × 3 AUC matrices of the five baseline models (mean ± SD over three seeds). The diagonal is in-domain. All models are close to perfect on ACRIMA (0.996 to 1.000), good on RIM-ONE DL (0.924 to 0.940) and clearly weaker on ORIGA (0.750 to 0.846). Off the diagonal, the AUC falls to between 0.506 and 0.811.

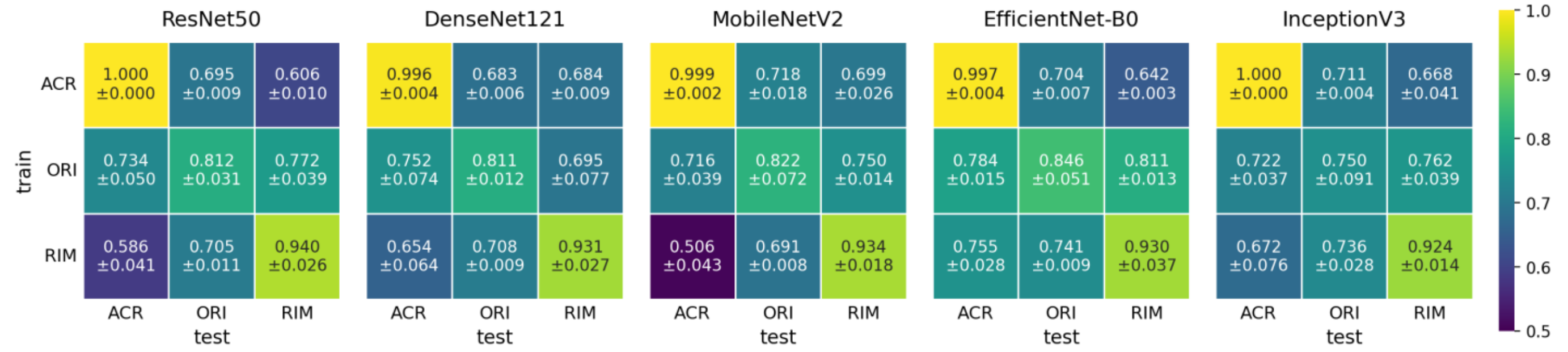

Fig. 3. Mean AUC ± SD over seeds 42, 7 and 123 for each model. Rows: training dataset; columns: test dataset (ACR: ACRIMA, ORI: ORIGA, RIM: RIM-ONE DL). The diagonal is in-domain.

Table II summarizes the models (Fig. 4). The in-domain AUC was between 0.891 (InceptionV3) and 0.924 (EfficientNet-B0). The cross-source AUC was between 0.680 (MobileNetV2) and 0.739 (EfficientNet-B0). Averaged over the five models, the in-domain AUC was 0.913 and the cross-source AUC was 0.702. The AUC drop was 0.211 ± 0.030. Every model lost at least 0.179. MobileNetV2 had the largest drop (0.238 ± 0.040), followed closely by ResNet50 (0.234 ± 0.021). InceptionV3 had the smallest drop (0.179 ± 0.058). The small drop of InceptionV3 comes mainly from its low in-domain AUC on ORIGA (0.750). It does not come from better transfer. Its cross-source AUC (0.712) is close to that of DenseNet121 (0.696) and lower than that of EfficientNet-B0 (0.739).

The 95% confidence interval of the average drop is 0.135 to 0.286. The interval of every single model excludes zero. The narrowest lower bound is 0.035, for InceptionV3. Paired t-tests on the seed-level values gave an uncorrected p below 0.05 for only two pairs, both for cross-source AUC: MobileNetV2 against EfficientNet-B0 ($p = 0.021$) and ResNet50 against EfficientNet-B0 ($p = 0.043$). For the drop itself, no pair reached $p < 0.05$ (the smallest were $p = 0.055$ for MobileNetV2 against EfficientNet-B0 and $p = 0.064$ for MobileNetV2 against InceptionV3). None of them stays below the Bonferroni threshold of 0.005 for the ten model pairs. We treat them as hints and do not claim a ranking of the architectures. The data support only that no architecture escaped the drop.

TABLE II
AUC of Each Model, Mean ± SD over Three Seeds

| Model | In-domain | Cross-source | AUC drop |
|---|---|---|---|
| ResNet50 | 0.917 ± 0.009 | 0.683 ± 0.019 | 0.234 ± 0.021 |
| DenseNet121 | 0.913 ± 0.009 | 0.696 ± 0.034 | 0.217 ± 0.038 |
| MobileNetV2 | 0.918 ± 0.024 | 0.680 ± 0.018 | 0.238 ± 0.040 |
| EfficientNet-B0 | 0.924 ± 0.015 | 0.739 ± 0.002 | 0.185 ± 0.017 |
| InceptionV3 | 0.891 ± 0.033 | 0.712 ± 0.026 | 0.179 ± 0.058 |
| **All models** | **0.913 ± 0.018** | **0.702 ± 0.014** | **0.211 ± 0.030** |

The last row is the mean of the five models computed for each seed. Its SD is taken over seeds.

## B. Dependence on the Source and Target Dataset

The size of the drop depends strongly on the source dataset (Table III, mean over five models and three seeds). Models trained on ACRIMA reached 0.998 in-domain but only 0.681 on the other two datasets, a drop of 0.317 ± 0.010. For RIM-ONE DL the drop was 0.256 ± 0.038. For ORIGA it was only 0.058 ± 0.056, but this is mostly because the in-domain AUC on ORIGA is low (0.808 ± 0.048). ORIGA-trained models gave the best cross-source AUC (0.750). This may relate to our own 2.5× disc-centered crop of ORIGA, to its lower glaucoma prevalence (26% versus 56% in ACRIMA), or to its harder in-domain task, so we do not attribute it to the dataset alone. In 6 of the 90 cross-source cells (5 models × 3 seeds × 6 source-target pairs), the cross-source AUC was higher than the in-domain AUC. All six belong to ORIGA-trained models. The ORIGA test split has 98 images with about 25 glaucoma cases, so its in-domain AUC is a noisy estimate.

TABLE III
Mean AUC over Five Models and Three Seeds

| Train \ Test | ACRIMA | ORIGA | RIM-ONE DL | Cross-source |
|---|---|---|---|---|
| ACRIMA | 0.998 ± 0.002 | 0.702 ± 0.007 | 0.660 ± 0.014 | 0.681 ± 0.010 |
| ORIGA | 0.742 ± 0.024 | 0.808 ± 0.048 | 0.758 ± 0.019 | 0.750 ± 0.020 |
| RIM-ONE DL | 0.635 ± 0.028 | 0.716 ± 0.007 | 0.932 ± 0.023 | 0.675 ± 0.016 |

The target dataset matters much less. The mean cross-source AUC was 0.688 when ACRIMA was the target, 0.709 for ORIGA and 0.709 for RIM-ONE DL. No dataset is a particularly hard target. The loss comes from the combination of source and target. For 37 of the 45 cells, the seed-to-seed SD is at most 0.05. It reaches 0.091 for InceptionV3 in-domain on ORIGA, 0.077 for DenseNet121 trained on ORIGA and tested on RIM-ONE DL, and 0.076 for InceptionV3 trained on RIM-ONE DL and tested on ACRIMA. The lowest cell mean is 0.506 (MobileNetV2 trained on RIM-ONE DL and tested on ACRIMA). With only three seeds, these SD values are themselves rough.

ACRIMA is almost perfect in-domain, so we checked how much it drives the average. Without ACRIMA as a source, the in-domain AUC is 0.870, the cross-source AUC is 0.713 and the drop is 0.157 ± 0.042 (95% CI 0.054 to 0.261). Without ACRIMA as both source and target, only ORIGA and RIM-ONE DL remain. The drop is then 0.133 ± 0.032 (95% CI 0.053 to 0.212) and the cross-source AUC is 0.737. The drop is smaller without ACRIMA, but it is still there and its confidence interval excludes zero.

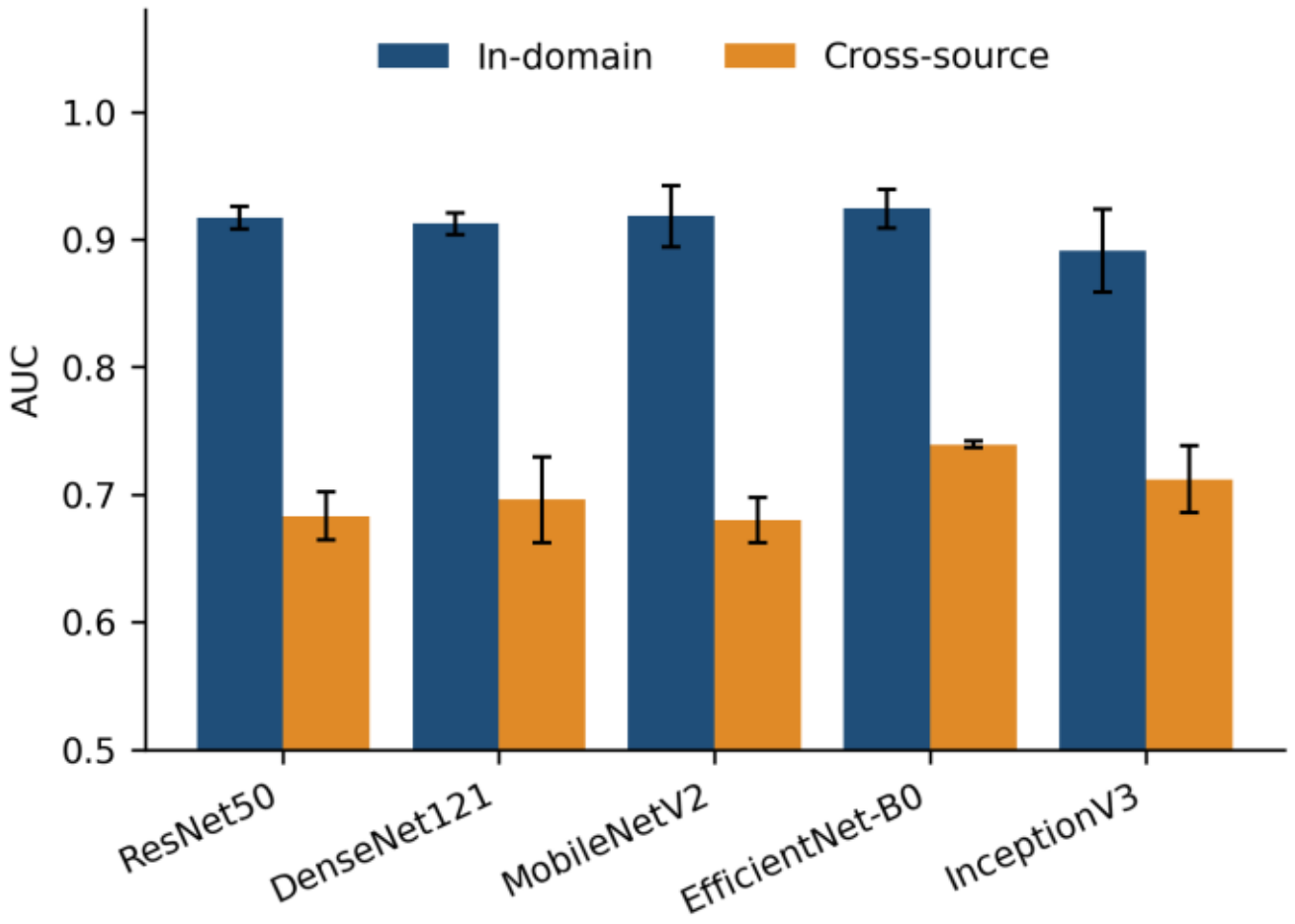

Fig. 4. In-domain and cross-source AUC of each model (mean ± SD over three seeds).

## C. Sensitivity and Specificity at the Source Threshold

Table IV gives the results at the threshold chosen on the source validation set. Averaged over the five models, sensitivity went from 0.831 in-domain to 0.726 cross-source. Specificity went from 0.870 to 0.506, and accuracy from 0.859 to 0.593. Specificity suffers most. A clear case is ACRIMA as the source and RIM-ONE DL as the target. All five models had sensitivity between 0.89 and 0.95 but specificity between 0.10 and 0.25. This means they called most RIM-ONE DL images glaucoma. The shift does not always go in this direction. DenseNet121 trained on ACRIMA and tested on ORIGA had sensitivity 0.492 and specificity 0.754. In both cases the threshold that was best on the source validation data is not best on the new data. The source threshold is also poorly determined. In 13 of the 15 ACRIMA-source networks the validation AUC was exactly 1.0, so many cutoffs are equally good on that validation set. Over all 45 networks the chosen thresholds ranged from 0.08 to 0.95. Part of the specificity loss may therefore come from an arbitrary cutoff and not only from dataset shift. Without ACRIMA as a source, the mean cross-source specificity was 0.570 ± 0.021 (0.506 with all sources), so the loss remains.

TABLE IV
Sensitivity, Specificity and Accuracy (Mean ± SD over Seeds)

| Model | Setting | Sensitivity | Specificity | Accuracy |
|---|---|---|---|---|
| ResNet50 | In-domain | 0.871 ± 0.032 | 0.836 ± 0.040 | 0.847 ± 0.023 |
| | Cross-source | 0.780 ± 0.054 | 0.458 ± 0.076 | 0.592 ± 0.016 |

| Model | Setting | Sensitivity | Specificity | Accuracy |
|---|---|---|---|---|
| DenseNet121 | In-domain | 0.827 ± 0.044 | 0.863 ± 0.026 | 0.857 ± 0.006 |
| | Cross-source | 0.723 ± 0.026 | 0.489 ± 0.036 | 0.585 ± 0.036 |
| MobileNetV2 | In-domain | 0.841 ± 0.074 | 0.887 ± 0.057 | 0.873 ± 0.016 |
| | Cross-source | 0.728 ± 0.078 | 0.463 ± 0.104 | 0.569 ± 0.053 |
| EfficientNet-B0 | In-domain | 0.847 ± 0.068 | 0.865 ± 0.057 | 0.857 ± 0.017 |
| | Cross-source | 0.774 ± 0.097 | 0.491 ± 0.170 | 0.611 ± 0.061 |
| InceptionV3 | In-domain | 0.767 ± 0.027 | 0.901 ± 0.011 | 0.862 ± 0.012 |
| | Cross-source | 0.626 ± 0.071 | 0.630 ± 0.119 | 0.607 ± 0.042 |

### *D. External Test on G1020*

The vCDR values of G1020 hardly separate the two classes (Fig. 5). The mean vCDR was 0.461 (SD 0.098) for normal and 0.469 (SD 0.097) for glaucoma images. The medians were 0.453 and 0.465. Used as a classifier, the vCDR gave an AUC of only 0.526. The horizontal ratio gave 0.536 and the area ratio 0.533. In a set with clinical glaucoma labels, one expects a clear shift of the cup-to-disc ratio between the classes. For this reason, we did not use G1020 as a source in the main experiments, and we used it only as an external test set.

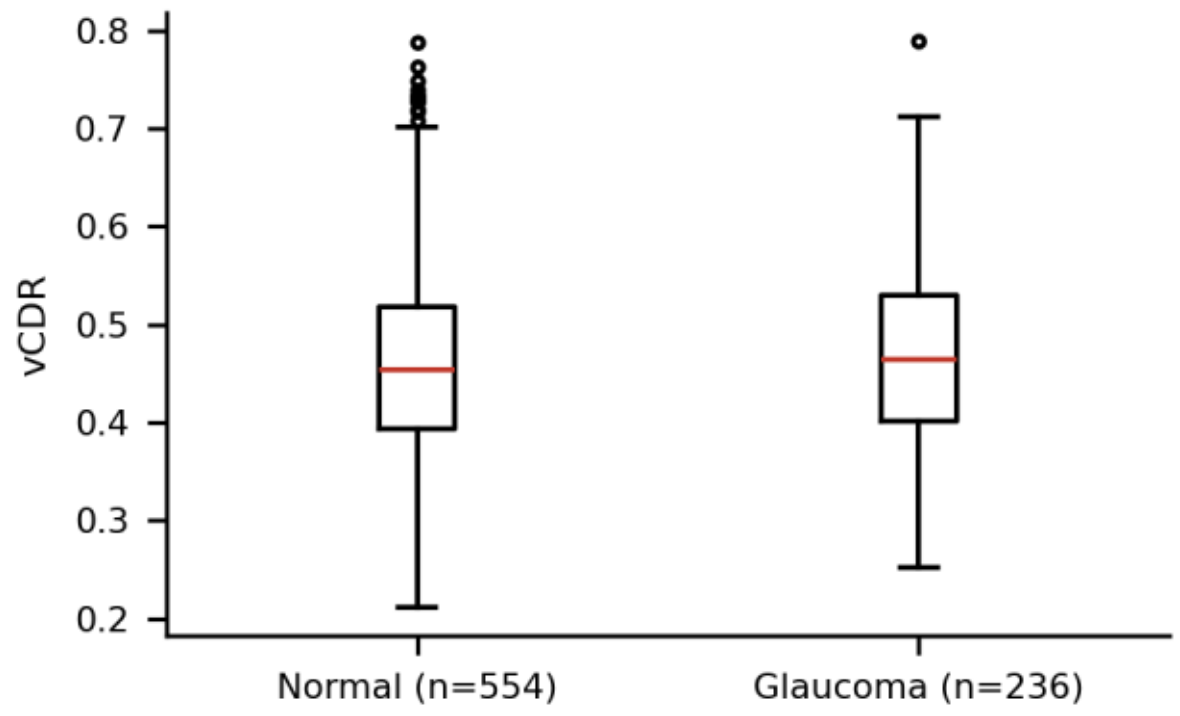

Fig. 5. G1020: vertical cup-to-disc ratio by class (790 images with measurable disc and cup).

The baseline models agreed with this finding (Table V). On G1020, the mean AUC of every model-source pair was between 0.472 and 0.511. The 45 single evaluations (5 models, 3 sources, 3 seeds) ranged from 0.458 to 0.529, and their mean was 0.493. This is chance level, even for the ORIGA-trained models that reached 0.75 on the two other external datasets. The external drop (source in-domain AUC minus G1020 AUC) was between 0.26 and 0.53. It mostly shows the fall to chance. At the stored thresholds, ACRIMA-trained models called most G1020 images glaucoma (sensitivity 0.855, specificity 0.126). ORIGA- and RIM-ONE DL-trained models had sensitivity of 0.344 and 0.389 and specificity of 0.661 and 0.612. At an AUC of 0.5, these values only show where the threshold lies. They do not show any ability to detect glaucoma.

TABLE V
G1020 AUC BY TRAINING SOURCE (MEAN ± SD OVER THREE SEEDS)

| Model | ACRIMA | ORIGA | RIM-ONE DL |
|---|---|---|---|
| ResNet50 | 0.486 ± 0.003 | 0.500 ± 0.006 | 0.511 ± 0.007 |
| DenseNet121 | 0.484 ± 0.013 | 0.497 ± 0.009 | 0.508 ± 0.003 |
| MobileNetV2 | 0.477 ± 0.006 | 0.509 ± 0.027 | 0.496 ± 0.006 |
| EfficientNet-B0 | 0.472 ± 0.008 | 0.498 ± 0.010 | 0.493 ± 0.006 |
| InceptionV3 | 0.494 ± 0.007 | 0.488 ± 0.026 | 0.485 ± 0.002 |

As a further check, the same five networks trained and tested inside G1020 (Section III-F) also stayed near chance. The mean test AUC was 0.523 ± 0.044 over models and seeds. The 15 single runs ranged from 0.433 to 0.617, with 44 glaucoma images in the test part.

### *E. Grad-CAM*

Fig. 6 shows Grad-CAM maps of the ACRIMA-trained ResNet50 on eight in-domain test images. For the five glaucoma images, the strongest activation is on the optic disc and its inner part. In one image the activation is wider and has two peaks. For the three normal images, the maps show little or no activation. ACRIMA images are already disc crops, so attention in the center is not surprising. The maps do not prove that the network uses the cup and rim the way a doctor does. We did not compute maps for cross-source predictions, so we cannot say whether the attention moves on foreign data.

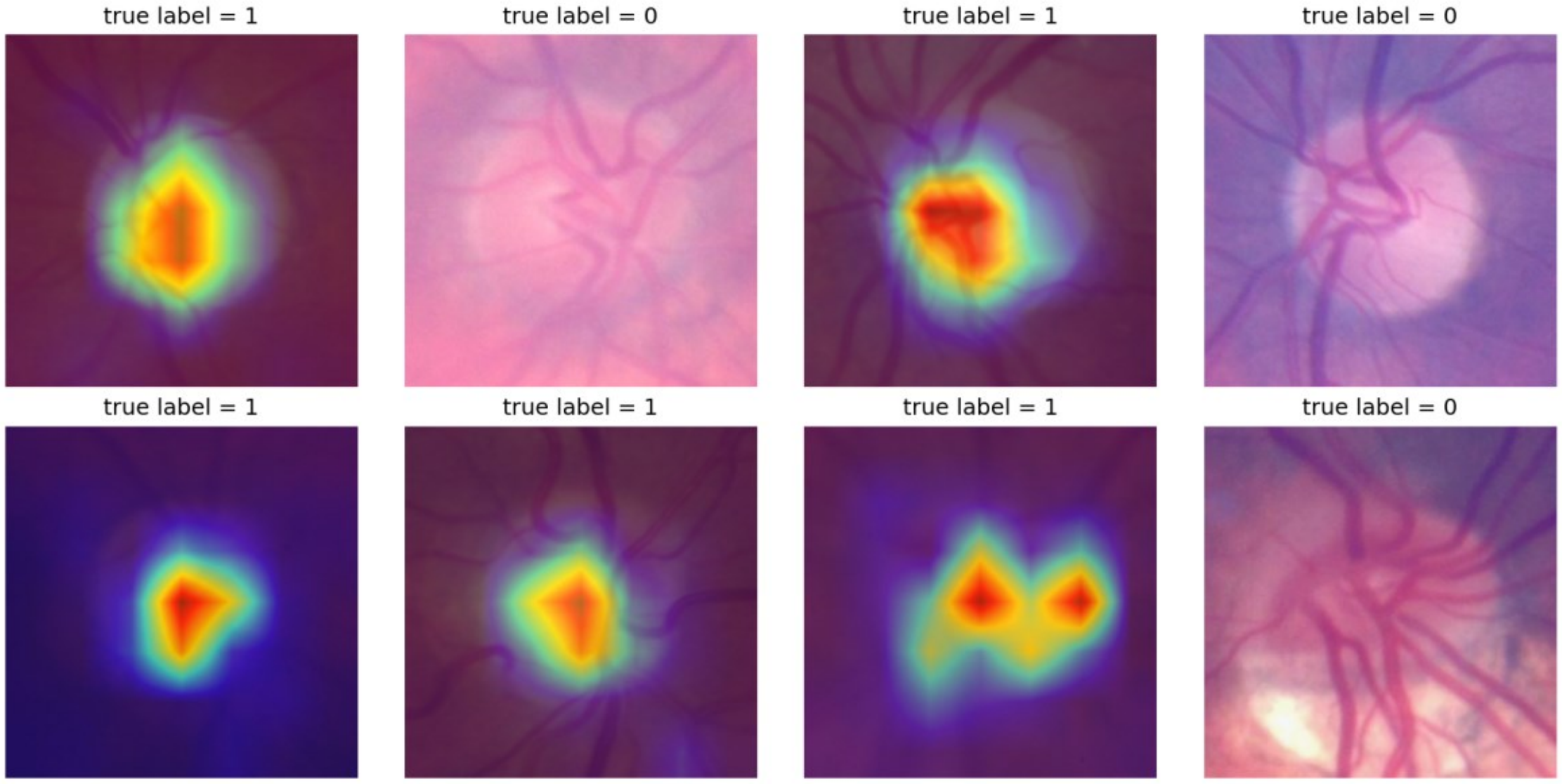

Fig. 6. Grad-CAM of ResNet50 trained on ACRIMA, on eight test images with their true labels.

### *F. Threshold Recalibration*

Table VI and Fig. 7 compare the source threshold with the threshold set again on 10% of the target images and tested on the other 90%. Averaged over the five models, recalibration raised specificity from 0.506 to 0.670 (change +0.164 ± 0.038; 95% CI 0.069 to 0.259). It lowered sensitivity from 0.726 to 0.609 (change −0.117 ± 0.009; 95% CI −0.139 to −0.095). The new threshold moved the operating point from many false positives to a more even balance of the two errors. It did not improve both. Accuracy rose from 0.593 to 0.643 (+0.051 ± 0.027) and balanced accuracy rose from 0.616 to 0.639 (+0.023 ± 0.015). With three seeds, the confidence intervals of both changes include zero (−0.016 to 0.117 and −0.014 to 0.061). All five models showed the same pattern of higher specificity and lower sensitivity. Balanced accuracy was higher for all five, from +0.007 for ResNet50 to +0.033 for EfficientNet-B0. InceptionV3 had the highest specificity at the source threshold (0.630). It gained the least specificity (+0.054) and lost almost no sensitivity (−0.015).

TABLE VI
CROSS-SOURCE PERFORMANCE BEFORE AND AFTER RECALIBRATION ON 10% OF THE TARGET IMAGES (MEAN ± SD OVER SEEDS)

| Model | Setting | Sensitivity | Specificity | Accuracy | Bal. acc. |
|---|---|---|---|---|---|
| ResNet50 | Source | 0.780 ± 0.054 | 0.458 ± 0.076 | 0.592 ± 0.016 | 0.619 ± 0.013 |
| | Recal. | 0.619 ± 0.029 | 0.634 ± 0.024 | 0.622 ± 0.013 | 0.626 ± 0.013 |
| DenseNet121 | Source | 0.723 ± 0.026 | 0.489 ± 0.036 | 0.585 ± 0.036 | 0.606 ± 0.027 |
| | Recal. | 0.589 ± 0.020 | 0.673 ± 0.030 | 0.639 ± 0.025 | 0.631 ± 0.025 |
| MobileNetV2 | Source | 0.728 ± 0.078 | 0.463 ± 0.104 | 0.569 ± 0.053 | 0.595 ± 0.021 |
| | Recal. | 0.597 ± 0.014 | 0.658 ± 0.029 | 0.636 ± 0.009 | 0.627 ± 0.012 |
| EfficientNet-B0 | Source | 0.774 ± 0.097 | 0.490 ± 0.170 | 0.611 ± 0.060 | 0.632 ± 0.038 |
| | Recal. | 0.629 ± 0.001 | 0.701 ± 0.008 | 0.668 ± 0.006 | 0.665 ± 0.004 |
| InceptionV3 | Source | 0.626 ± 0.070 | 0.630 ± 0.119 | 0.606 ± 0.042 | 0.628 ± 0.025 |
| | Recal. | 0.611 ± 0.021 | 0.684 ± 0.036 | 0.651 ± 0.028 | 0.647 ± 0.023 |
| All models | Source | 0.726 ± 0.011 | 0.506 ± 0.022 | 0.593 ± 0.016 | 0.616 ± 0.007 |
| | Recal. | 0.609 ± 0.010 | 0.670 ± 0.017 | 0.643 ± 0.012 | 0.639 ± 0.010 |

Source: source threshold. Recal.: recalibrated threshold. Both are evaluated on the remaining 90% of the target images. Bal. acc.: balanced accuracy.

Recalibration also reduced the spread between seeds and between models. The SD of specificity over seeds fell from 0.170 to 0.008 for EfficientNet-B0 and from 0.104 to 0.029 for MobileNetV2. This is expected, because the threshold is now set on target images and not on the source validation set. By source and target pair, balanced accuracy increased in all six pairs, from +0.008 (ACRIMA to ORIGA) to +0.052 (ACRIMA to RIM-ONE DL). The latter is the case where the source threshold had called most images glaucoma. There, recalibration changed specificity from 0.187 to 0.602 and sensitivity from 0.917 to 0.606. Over the 30 combinations of model, source and target (mean over seeds), balanced accuracy increased in 23 and decreased in 7.

On G1020, recalibration did not help. Sensitivity went from 0.529 to 0.476, specificity from 0.466 to 0.516, and balanced accuracy from 0.498 to 0.496 (change −0.002 ± 0.002). This is expected at an AUC of 0.5. A threshold on a score that does not rank the images can only move the operating point along the chance line. The size of the calibration sample mattered little. With 5%, 10%, 20% and 30% of the target images, the mean balanced accuracy was 0.634, 0.639, 0.644 and 0.645, against 0.616 at the source threshold. From 5% to 30% calibration data, the mean sensitivity rose from 0.587 to 0.625, and the mean specificity fell from 0.680 to 0.665.

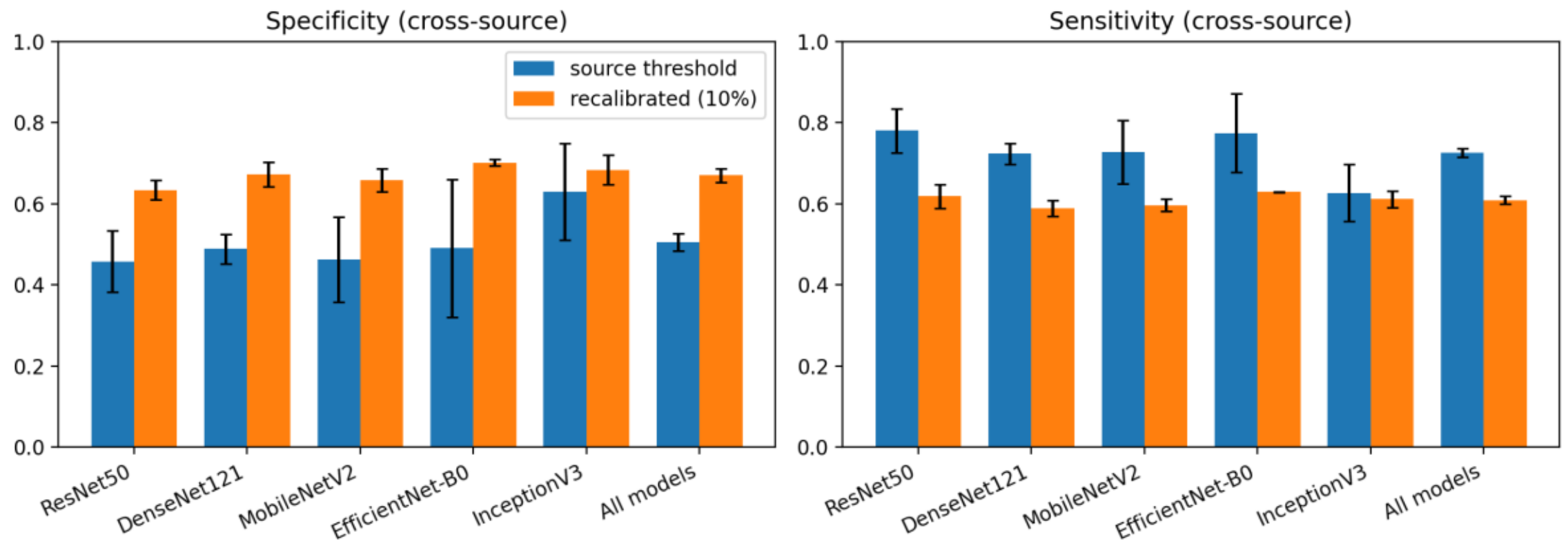

Fig. 7. Cross-source specificity and sensitivity with the source threshold and with the threshold recalibrated on 10% of the target images (mean ± SD over three seeds, evaluated on the remaining 90%).

### *G. Multi-Source versus Single-Source Training*

Table VII and Fig. 8 compare networks trained on two datasets together with single-source networks tested on the same held-out dataset. Averaged over the five models, the AUC on the held-out dataset was 0.695 ± 0.023 after multi-source training. For the mean of the two single-source models it was 0.702 ± 0.014. The difference is −0.007 (SD 0.012; 95% CI −0.036 to 0.023). Multi-source training therefore did not improve generalization. The rule we fixed in advance (a confidence interval that excludes zero and a gain of at least 0.02) was not met. The sign of the gain differed between models, from −0.020 (EfficientNet-B0) to +0.013 (ResNet50). Compared with the better of the two single sources, the multi-source AUC was lower by 0.048 (95% CI −0.092 to −0.003). One cannot pick the better source without labels from the target. Compared with the worse one, the multi-source AUC was higher by 0.034 (95% CI 0.019 to 0.049). Pooling protected against a poor choice of source, but it did not beat a good one.

The effect depended on the held-out dataset (Table VIII). For ACRIMA and ORIGA the gain was small (+0.009 and +0.008). For RIM-ONE DL the AUC fell by 0.038, from 0.709 to 0.671. The three targets are not independent, because each dataset is in two of the pools. The size-matched control gave a held-out AUC of 0.712 ± 0.014 and a gain of +0.010 (SD 0.006; 95% CI −0.004 to 0.025). This also does not meet the rule. Its gains by target were +0.043 (ACRIMA), +0.031 (ORIGA) and −0.043 (RIM-ONE DL). The size-matched AUC is higher than that of the pool with all images (0.695). The difference is within the seed variation and was not tested. We conclude only that neither arm gave a reliable gain.

TABLE VII
AUC ON THE HELD-OUT DATASET, SINGLE-SOURCE VERSUS MULTI-SOURCE TRAINING, BY MODEL (MEAN ± SD OVER THREE SEEDS)

| Model | Single-source (mean of 2) | Best single | Multi-source | Gain | Size-matched | Gain |
|---|---|---|---|---|---|---|
| ResNet50 | 0.683 ± 0.019 | 0.738 ± 0.022 | 0.696 ± 0.037 | +0.013 ± 0.040 | 0.704 ± 0.025 | +0.021 ± 0.017 |
| DenseNet121 | 0.696 ± 0.034 | 0.724 ± 0.044 | 0.682 ± 0.031 | −0.014 ± 0.007 | 0.709 ± 0.037 | +0.013 ± 0.021 |
| MobileNetV2 | 0.680 ± 0.018 | 0.728 ± 0.016 | 0.685 ± 0.035 | +0.005 ± 0.022 | 0.695 ± 0.019 | +0.015 ± 0.012 |
| EfficientNet-B0 | 0.739 ± 0.002 | 0.780 ± 0.005 | 0.720 ± 0.019 | −0.020 ± 0.018 | 0.746 ± 0.022 | +0.007 ± 0.023 |
| InceptionV3 | 0.712 ± 0.026 | 0.745 ± 0.027 | 0.693 ± 0.024 | −0.019 ± 0.003 | 0.708 ± 0.024 | −0.004 ± 0.006 |
| **All models** | **0.702 ± 0.014** | **0.743 ± 0.014** | **0.695 ± 0.023** | **−0.007 ± 0.012** | **0.712 ± 0.014** | **+0.010 ± 0.006** |

Single-source: the two networks trained on each of the other two datasets (mean and better of the two). Multi-source: trained on the two datasets together. Size-matched: pool as large as the smaller single-source training set. Gain: change from the mean of the two single-source AUCs.

TABLE VIII
AUC ON THE HELD-OUT DATASET BY HELD-OUT TARGET (MEAN OVER FIVE MODELS, SD OVER SEEDS)

| Held-out target | Single-source (mean of 2) | Best single | Multi-source | Gain | Size-matched | Gain |
|---|---|---|---|---|---|---|
| ACRIMA | 0.688 ± 0.026 | 0.745 ± 0.024 | 0.697 ± 0.045 | +0.009 ± 0.025 | 0.731 ± 0.049 | +0.043 ± 0.025 |
| ORIGA | 0.709 ± 0.006 | 0.722 ± 0.008 | 0.717 ± 0.007 | +0.008 ± 0.003 | 0.741 ± 0.017 | +0.031 ± 0.022 |
| RIM-ONE DL | 0.709 ± 0.011 | 0.761 ± 0.014 | 0.671 ± 0.020 | −0.038 ± 0.013 | 0.665 ± 0.021 | −0.043 ± 0.022 |

Columns as in Table VII.

The AUC drop for multi-source networks was defined as the in-domain AUC on the source test parts minus the held-out AUC. It was 0.218 ± 0.040 (in-domain 0.913, held-out 0.695), against 0.211 ± 0.030 for single-source networks (change +0.008; 95% CI −0.018 to 0.033). In the size-matched arm the drop was 0.183 ± 0.030 (change −0.028; 95% CI −0.031 to −0.025). This came mostly from a lower in-domain AUC (0.895) and less from a higher held-out AUC (0.712). The interval is narrow because the SD over three seeds is only 0.001, so it should not be over-read.

At the threshold chosen on the pooled validation sets, the mean specificity on the held-out dataset was 0.467 ± 0.041 (single-source 0.506 ± 0.022). The mean sensitivity was 0.757 ± 0.051 (single-source 0.726 ± 0.011) and the balanced accuracy was 0.612 (single-source 0.616). For RIM-ONE DL the specificity was 0.245 (single-source 0.416) and the sensitivity was 0.916 (single-source 0.798). The ranking did not improve, and neither did the operating point. When we recalibrated the multi-source networks on 10% of the held-out dataset (Section III-H), the mean specificity rose from 0.467 to 0.654 and the mean sensitivity fell from 0.757 to 0.615. Balanced accuracy rose from 0.612 to 0.635 (+0.023; 95% CI −0.015 to 0.060). This is the same gain as for single-source networks (Section IV-F). On G1020 the mean AUC was 0.495 for multi-source networks (0.494 in the size-matched arm) and 0.493 for single-source networks. Pooling did not change the chance-level result.

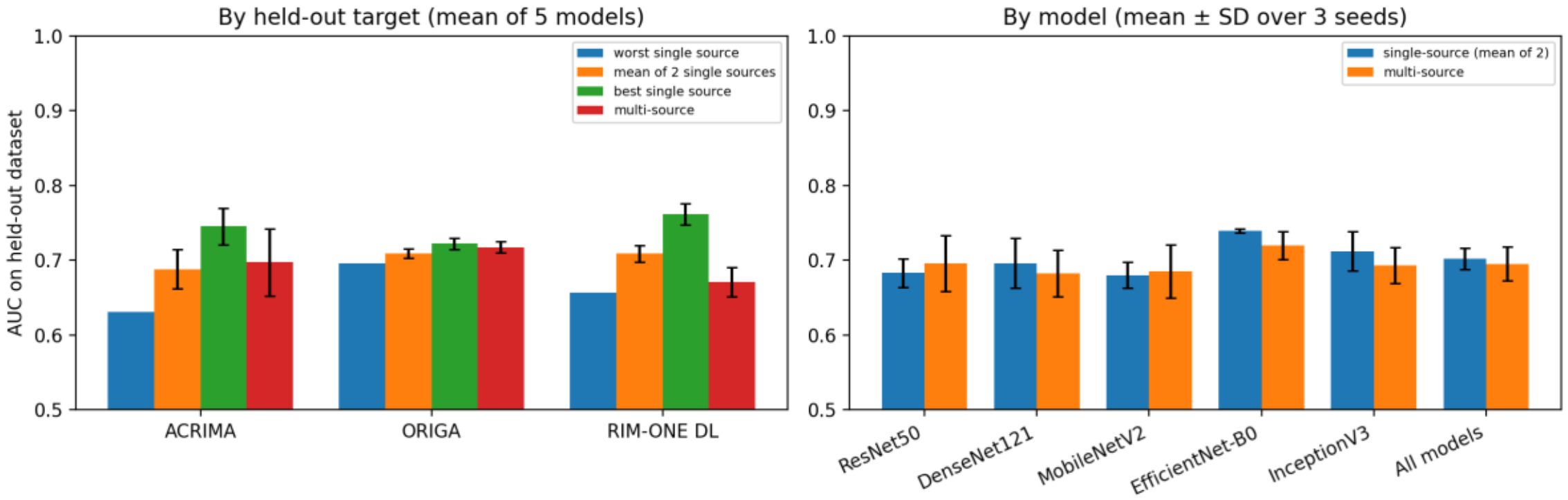

Fig. 8. AUC on the held-out dataset for single-source models (worst, mean and best of the two sources) and for multi-source models trained on the other two datasets. Left: by held-out target, mean of five models. Right: by model. Error bars: SD over seeds 42, 7 and 123.

### *H. Normalization and Color Augmentation*

Table IX and Fig. 9 show that the effect depended on the method. LAB reference normalization had the highest all-model mean cross-source AUC, 0.755 ± 0.011, against 0.702 ± 0.014 for the baseline, but strong color augmentation was nearly identical (0.754), and this best condition was identified only after seeing the results. The paired gain was 0.053 (95% CI 0.046 to 0.060), and the in-domain AUC changed only from 0.913 to 0.910. The drop went from 0.211 to 0.155 ± 0.028. The paired reduction was 0.056 (95% CI 0.048 to 0.063), which is about 26% of the baseline drop. Strong color augmentation reached a cross-source AUC of 0.754 ± 0.007, a gain of 0.052 (95% CI 0.033 to 0.071), with a remaining drop of 0.158 ± 0.019. The two averages are close, but this does not show that the two methods are equal or give a reliable ranking.

Ben Graham normalization raised the cross-source AUC to 0.733 ± 0.009, a paired gain of 0.031 (95% CI 0.019 to 0.043), and reduced the drop to 0.177 ± 0.027. CLAHE gave a smaller gain of 0.010 (95% CI 0.005 to 0.014) and reached 0.712 ± 0.014. Gray-world normalization did not improve the mean transfer. Its cross-source AUC was 0.696 ± 0.008, a change of −0.006 (95% CI −0.023 to 0.011), although its in-domain AUC went up to 0.920. The unadjusted p-values of the cross-source gains were 0.0010 for LAB reference, 0.0073 for strong color augmentation, 0.0078 for Ben Graham and 0.0142 for CLAHE. The first three are below the threshold of 0.01 for five comparisons. CLAHE is not. These are exploratory comparisons based on three seeds.

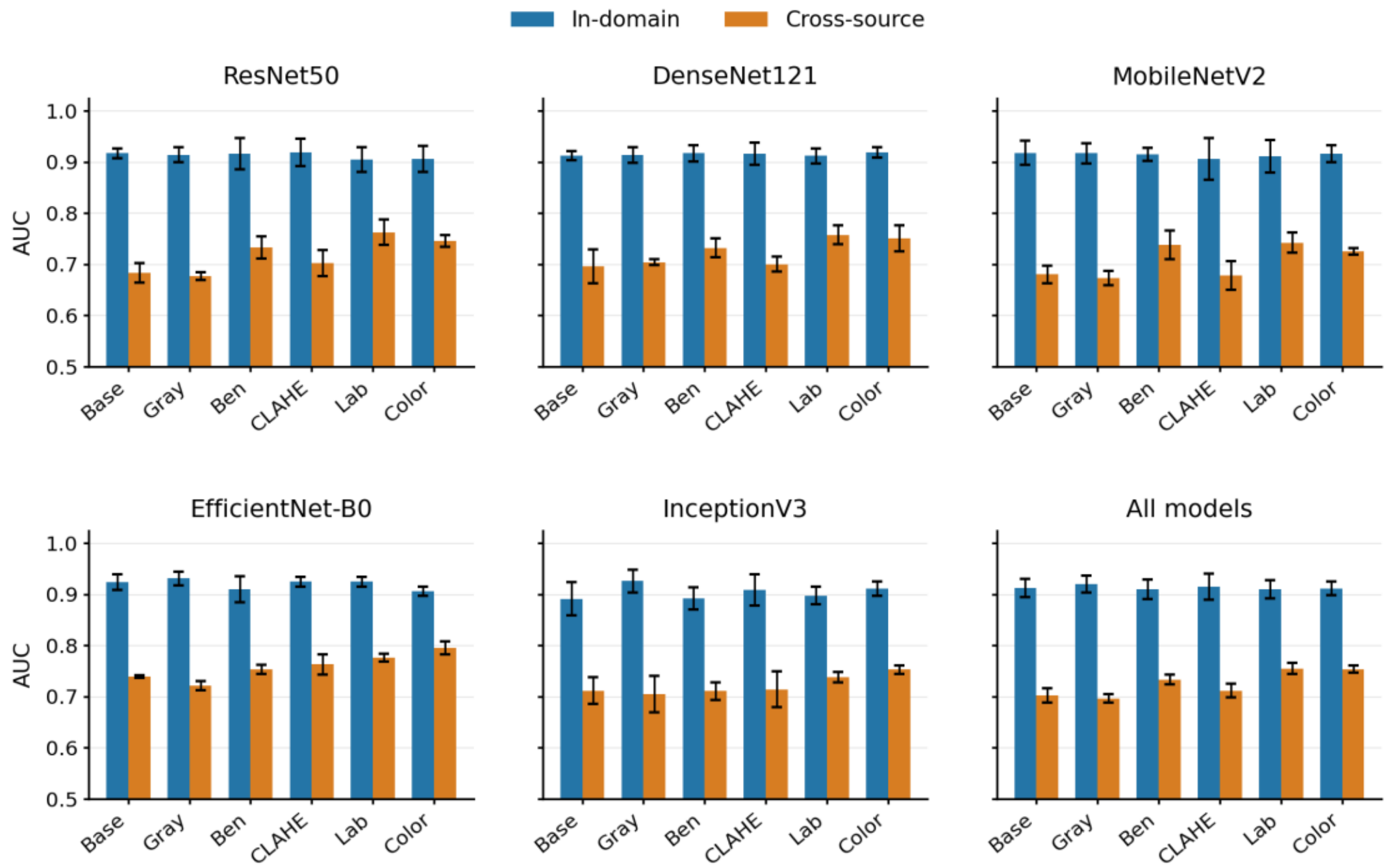

Fig. 9. In-domain and cross-source AUC under each appearance condition, by architecture and averaged over models. Error bars show SD over three seeds. Base: baseline; Gray: gray-world; Ben: Ben Graham; Lab: LAB reference; Color: strong color augmentation. The vertical axis starts at 0.5.

TABLE IX
NORMALIZATION AND AUGMENTATION EFFECTS ON AUC, AVERAGED OVER ALL MODELS

| Condition | In-domain AUC | Cross-source AUC | AUC drop | Cross-source gain [95% CI] | p-value |
|---|---|---|---|---|---|
| Baseline | 0.913 ± 0.018 | 0.702 ± 0.014 | 0.211 ± 0.030 | – | – |
| Gray-world | 0.920 ± 0.016 | 0.696 ± 0.008 | 0.224 ± 0.023 | −0.006 [−0.023, +0.011] | 0.2781 |
| Ben Graham | 0.910 ± 0.019 | 0.733 ± 0.009 | 0.177 ± 0.027 | +0.031 [+0.019, +0.043] | 0.0078 |
| CLAHE | 0.915 ± 0.025 | 0.712 ± 0.014 | 0.203 ± 0.037 | +0.010 [+0.005, +0.014] | 0.0142 |
| LAB reference | 0.910 ± 0.018 | 0.755 ± 0.011 | 0.155 ± 0.028 | +0.053 [+0.046, +0.060] | 0.0010 |
| Strong color aug. | 0.912 ± 0.013 | 0.754 ± 0.007 | 0.158 ± 0.019 | +0.052 [+0.033, +0.071] | 0.0073 |

AUC values and drops are mean ± SD over three seeds, after averaging the five models within each seed. The gain is paired with the baseline. Confidence intervals and p-values are not adjusted. The exploratory threshold for the five cross-source contrasts is 0.01. AUC drop = in-domain AUC minus cross-source AUC.

The response also depended on the architecture (Table X). LAB reference normalization and strong color augmentation raised the mean cross-source AUC for all five models. The largest model-level gain with LAB reference was for ResNet50, from 0.683 to 0.763. EfficientNet-B0 with strong color augmentation had the highest cross-source AUC of all model and condition pairs (0.795 ± 0.013). We do not claim it is the best, because we looked at many combinations. Ben Graham normalization raised four of the five architecture means, while InceptionV3 stayed near its baseline (0.711 against 0.712). Gray-world normalization raised only DenseNet121. In-domain AUC changed little for most model-condition pairs, with two exceptions. Gray-world raised the in-domain AUC of InceptionV3 from 0.891 to 0.926 (+0.035; its baseline was the lowest of the five models), which explains most of the rise of the all-model mean to 0.920. Strong color augmentation lowered the in-domain AUC of EfficientNet-B0 from 0.924 to 0.906 (−0.018), even though this model had the highest cross-source AUC.

TABLE X
CROSS-SOURCE AUC BY ARCHITECTURE AND APPEARANCE CONDITION (MEAN ± SD OVER THREE SEEDS)

| Model | Baseline | Gray-world | Ben Graham | CLAHE | LAB reference | Strong color aug. |
|---|---|---|---|---|---|---|
| ResNet50 | 0.683 ± 0.019 | 0.677 ± 0.008 | 0.733 ± 0.021 | 0.702 ± 0.025 | 0.763 ± 0.025 | 0.746 ± 0.011 |
| DenseNet121 | 0.696 ± 0.034 | 0.704 ± 0.006 | 0.732 ± 0.018 | 0.700 ± 0.015 | 0.757 ± 0.019 | 0.751 ± 0.026 |
| MobileNetV2 | 0.680 ± 0.018 | 0.674 ± 0.014 | 0.738 ± 0.028 | 0.678 ± 0.028 | 0.742 ± 0.020 | 0.725 ± 0.007 |
| EfficientNet-B0 | 0.739 ± 0.002 | 0.722 ± 0.009 | 0.753 ± 0.009 | 0.763 ± 0.020 | 0.776 ± 0.008 | 0.795 ± 0.013 |
| InceptionV3 | 0.712 ± 0.026 | 0.705 ± 0.036 | 0.711 ± 0.017 | 0.714 ± 0.035 | 0.738 ± 0.010 | 0.753 ± 0.008 |

Each seed-level AUC is the mean of the six off-diagonal source-target cells. Differences between architectures are described only. We make no claim about which model is best.

Fig. 10 shows the changes by source and target. Of the 30 off-diagonal model-source-target cells, LAB reference normalization and strong color augmentation each raised 27 cell means. Ben Graham raised 24, CLAHE 20 and gray-world 15. These are counts, not independent tests. A large LAB gain was seen for RIM-ONE DL to ACRIMA: the mean over models went from 0.635 to 0.795. ORIGA to ACRIMA went from 0.742 to 0.801. With ACRIMA as the target, the mean LAB gain over the two off-diagonal cells was +0.110, against +0.016 with ORIGA and +0.033 with RIM-ONE DL as the target. Strong color augmentation raised ACRIMA to RIM-ONE DL from 0.660 to 0.745 and ORIGA to RIM-ONE DL from 0.758 to 0.822. Gray-world lowered the AUC of ACRIMA-trained models (−0.036 on ORIGA and −0.028 on RIM-ONE DL). The diagonal cells changed little. The changes of image appearance helped several hard transfers, but the size of the gain depended on the source and the target.

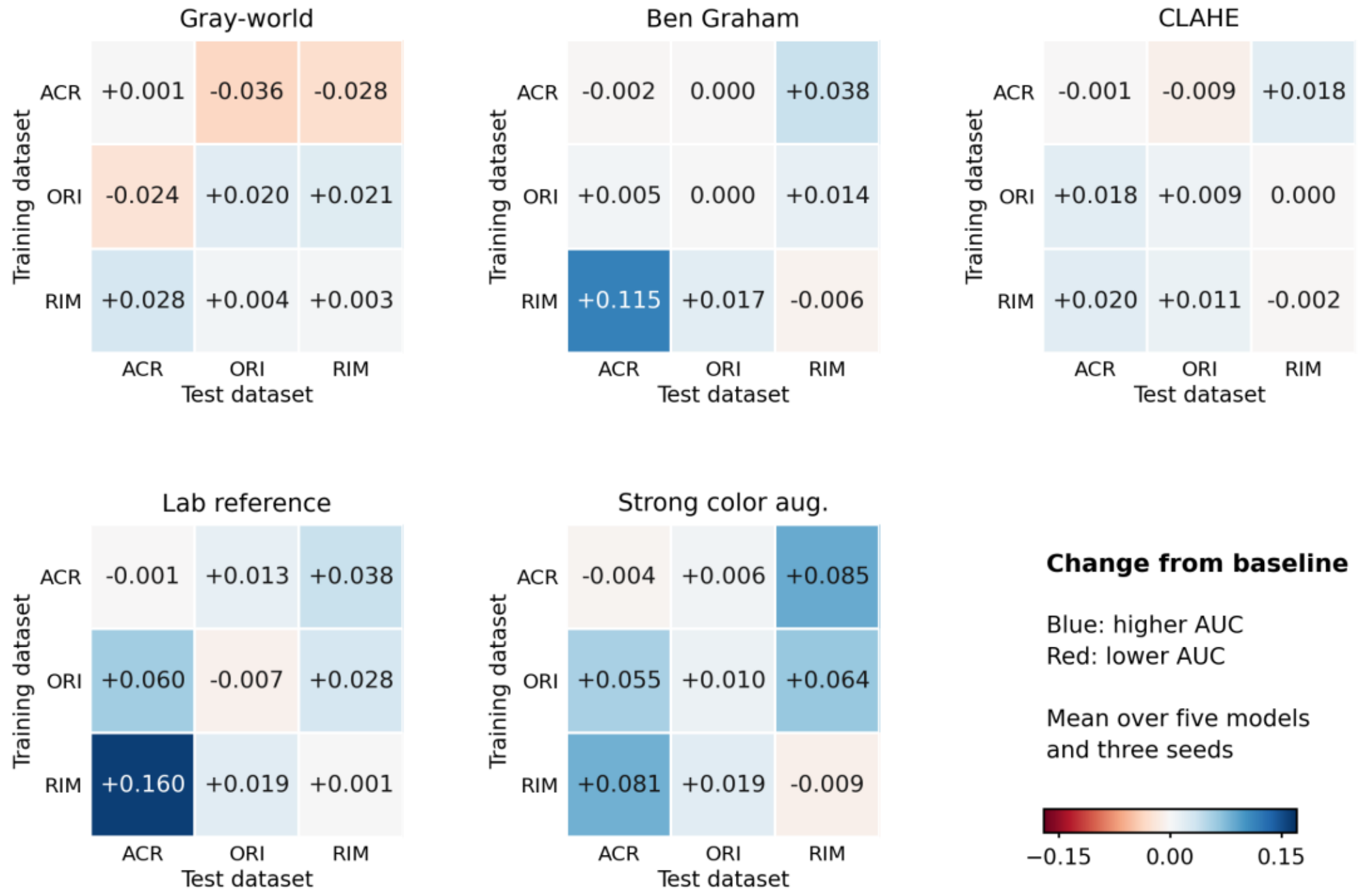

Fig. 10. Change in mean AUC relative to the baseline, averaged over five models and three seeds, for each normalization or augmentation condition. ACR: ACRIMA; ORI: ORIGA; RIM: RIM-ONE DL. Rows are training sources and columns are test targets. Diagonal cells are in-domain. Blue means an increase and red a decrease. Cell means show transfer patterns and are not independent statistical tests.

At the source-selected thresholds, a better AUC did not mean that sensitivity and specificity both improved (Table XI). LAB reference normalization raised the cross-source specificity from 0.506 to 0.674 and the accuracy from 0.593 to 0.666, but it lowered the sensitivity from 0.726 to 0.653. Strong color augmentation gave a specificity of 0.729, an accuracy of 0.673 and a sensitivity of 0.607. The balanced accuracy was 0.664 and 0.668, against 0.616 at baseline. Ben Graham normalization gave a sensitivity of 0.688 and a specificity of 0.615. These operating-point values are plain averages over models and cells. We did not estimate their uncertainty by averaging the model-level SD values.

TABLE XI
CROSS-SOURCE OPERATING-POINT METRICS AND SEPARATE G1020 AUC

| Condition | Sensitivity | Specificity | Accuracy | Balanced accuracy | G1020 AUC |
|---|---|---|---|---|---|
| Baseline | 0.726 | 0.506 | 0.593 | 0.616 | 0.493 |
| Gray-world | 0.715 | 0.507 | 0.592 | 0.611 | 0.496 |
| Ben Graham | 0.688 | 0.615 | 0.645 | 0.651 | 0.497 |
| CLAHE | 0.689 | 0.559 | 0.613 | 0.624 | 0.492 |
| LAB reference | 0.653 | 0.674 | 0.666 | 0.664 | 0.502 |
| Strong color aug. | 0.607 | 0.729 | 0.673 | 0.668 | 0.499 |

The first four metric columns are averages over the five models and the six cross-source cells at the source-selected thresholds. Balanced accuracy is the mean of sensitivity and specificity. G1020 AUC is a separate average over the three sources and five models. It is not part of the 3 × 3 average. The values summarize three seeds. No pooled SD or confidence interval is given.

The case of ACRIMA to RIM-ONE DL shows the trade clearly. In the baseline, the sensitivity was 0.917 and the specificity only 0.187. LAB reference normalization changed these to 0.653 and 0.637, with an AUC of 0.698. Strong color augmentation gave a sensitivity of 0.664, a specificity of 0.693 and an AUC of 0.745. Ben Graham gave an AUC of 0.698, a sensitivity of 0.775 and a specificity of 0.475. Gray-world made this transfer worse, with an AUC of 0.632 and a specificity of 0.139. The drop in false positives under the stronger conditions came with more false negatives at the source-selected threshold.

None of the changes brought back useful detection on G1020. The all-model mean AUC was 0.493 at baseline, 0.496 with gray-world, 0.497 with Ben Graham, 0.492 with CLAHE, 0.502 with LAB reference and 0.499 with strong color augmentation. LAB reference and strong augmentation moved the sensitivity and specificity from the baseline 0.529 and 0.466 to 0.418 and 0.587, and to 0.440 and 0.560, but the AUC stayed near 0.5. Changing the image appearance did not fix the G1020 problem in this pipeline.

## V. Discussion

In the baseline experiment, five standard CNNs scored 0.89 to 0.92 AUC on their own datasets and about 0.70 on other public datasets. An AUC of 0.70 means that a random glaucoma image gets a higher score than a random normal image in about seven of ten pairs. This is far from what a screening program needs, and far from what the in-domain numbers suggest. The threshold makes it worse. With a cutoff set on the source data, the average specificity on a new dataset was 0.51. This means that about half of the healthy eyes would be wrongly labeled.

In the baseline comparison, under these settings, no architecture protected against the drop. EfficientNet-B0 had the best cross-source AUC (0.739 ± 0.002). Its differences from DenseNet121 (0.696) and InceptionV3 (0.712) are within the seed variation. Its differences from ResNet50 and MobileNetV2 (uncorrected p = 0.043 and 0.021) do not survive the correction for multiple comparisons. Model size does not explain the order either. The small MobileNetV2 and the large ResNet50 had the two largest drops, while the compact EfficientNet-B0 had the best cross-source result. In this data, the source dataset accounted for more of the variation in AUC than the architecture did. This is descriptive only. The source effect is confounded with our crop choices, class prevalence and labeling rules, and we did not test it formally. The source dataset also determined the drop much more than the target did. ORIGA-trained models transferred best although they were the weakest in-domain, which may partly reflect the crop and prevalence differences noted above.

The normalization test looked at one possible cause of the drop. The datasets differ in color, illumination, field of view, image size and crop (Fig. 1). LAB reference normalization and strong color augmentation raised the mean cross-source AUC by about 0.05, and the in-domain AUC stayed near 0.91. This supports the idea that differences in appearance contribute to the transfer failure. It does not show that this is the only cause. Because strong color augmentation changes several factors at once (wider brightness and contrast ranges, plus saturation and hue), we cannot say which component produced the gain. An ablation of each factor is left for future work. The gains were largest when ACRIMA was the target, and ACRIMA has the most saturated orange-red images. This fits the color explanation, but we did not test it directly. The normalization results narrow the explanation, but they do not remove the problem. The best conditions reduced the drop by about one quarter, and a drop of about 0.16 remained. Gray-world balancing did not help on average and CLAHE gave only a small gain, so changing the appearance alone is not enough. The two conditions that helped most also moved the source-threshold operating point toward higher specificity and lower sensitivity. A better ranking must therefore be checked together with the errors that matter for screening. Balanced accuracy for LAB reference and color augmentation (0.664 and 0.668) was higher than after recalibration on 10% of the target images (0.639). We do not rank the two remedies by these numbers, because they come from different training runs and part of the change is a shift of the score distribution.

The datasets also come from different centers and populations, and their labels may follow different rules. A network can use any feature that separates its classes, including features that belong only to one dataset, such as the color of one camera or a typical crop [13], [14]. The ACRIMA result fits this idea. All models reached a validation AUC of at least 0.9996 on ACRIMA in all seeds, so the task is easy inside that dataset. The same models reached only 0.65 to 0.71 on the other two. This explains part of the average drop but not all of it. Without ACRIMA as a source the drop is 0.157, and with only ORIGA and RIM-ONE DL it is 0.133. The main finding therefore does not depend on one dataset. We stress that this is an interpretation. Our experiments do not separate image appearance from label differences. They also do not separate the dataset differences from our own crop choices: ORIGA and G1020 were cropped at 2.5 times the disc size, ACRIMA and RIM-ONE DL are tighter crops of varying size, and some images used automatic disc detection.

The G1020 result is a warning about label quality and about benchmarks in general. All 45 baseline external evaluations were at chance level. The cup-to-disc ratio computed from the masks of the dataset barely separated the classes (AUC 0.53). Networks trained and tested inside G1020 reached a mean test AUC of only 0.52. This last result matters. Even models fitted to G1020 stayed near chance, so the chance-level external AUC cannot be blamed on dataset shift alone. Either the G1020 labels depend on information that cannot be seen at the optic disc, or our pipeline cannot extract it from this dataset, or both. Our G1020 crops are looser than the disc crops of ACRIMA and RIM-ONE DL, so after resizing to 224 × 224 pixels the disc covers a smaller part of the input. We cannot tell which explanation is true. We did not train on the original uncropped G1020 images or compare our results with published G1020 results. Both would help to exclude an error in our crop or label handling, and we leave them for future work. We do not claim that the labels are wrong, and the test part of our in-domain check is small. The practical result is clear. If we had added G1020 to the training data, the models would have received labels that they could not link to the look of the disc. This is why we used it as a stress test and not as a fourth source. The mean external AUC stayed between 0.492 and 0.502 in all appearance conditions. This does not show that the labels are incorrect or that G1020 cannot be learned.

Simple pooling of two datasets did not solve the problem either. The held-out AUC of the multi-source networks (0.695) was not higher than that of the single-source networks (0.702), and it fell by 0.038 for RIM-ONE DL. The confidence interval of the difference (−0.036 to 0.023) fits a small loss or a small gain. With this design we can rule out a gain much larger than 0.02 AUC, but not a smaller one. Pooling did protect against a poor choice of source (+0.034 over the worse single source), but it did not beat the better one (−0.048). A pool of the size of one source gave a similar small gain that was not significant (+0.010), so more data was not clearly the cause in either direction. One possible reason, which we did not test, is that the two pooled datasets differ from each other as much as from the held-out one. A network that sees two appearances can learn the cues of both, and plain pooling gives it no reason to ignore them. The larger dataset can also dominate the training. This agrees with the earlier finding that the choice of training data mattered more than the architecture. It does not show that multi-source training can never help. We used three datasets, two sources per pool and three seeds. Normalization was tested separately and not together with pooling, and domain adaptation was not tested. Pooling also did not repair the threshold. The specificity at the pooled-validation threshold was 0.47, and recalibration on 10% of the target gave the same trade between sensitivity and specificity as for single-source networks.

The threshold chosen on the source validation set moved sensitivity and specificity in different directions, depending on the pair of datasets. The recalibration experiment (Section IV-F) shows both the use and the limit of setting it again. A labeled sample of 10% of the target images (48 to 70 images) partly corrected the main failure of the source threshold, the large number of false positives. Specificity rose from 0.51 to 0.67, and the variation between seeds and models became smaller. Sensitivity, however, fell from 0.73 to 0.61, and balanced accuracy rose by only 0.023, with a confidence interval that includes zero. This is what one expects at an AUC of about 0.70. A threshold moves the operating point along the ROC curve. It can correct a badly placed threshold, but it cannot create separation that the network does not have. On G1020, as expected at chance-level AUC, recalibration gave no gain.

Two practical points follow from this work. First, a model should be validated on data from the site where it will be used. The AUC on that data should be reported together with the in-domain value. Second, the operating threshold should be set again on a small labeled sample from the new site. A sensitivity of 0.61 would be too low for a screening program. At a screening prevalence of about 3.5% [1], a sensitivity of 0.61 and a specificity of 0.67 would also give a positive predictive value of only about 6%, so most referrals would be false alarms. Where sensitivity matters most, the threshold should be set for a target sensitivity and not with Youden's J. We did not test this. In our experiments, the normalization test improved the ranking more than plain pooling.

*Limitations*

*Data.* We used only three source datasets. All of them are small, and they contain more glaucoma cases than a real screening population would. The in-domain test sets are also small, with 73 to 106 images. ORIGA and RIM-ONE DL have only about 25 glaucoma images each, so in-domain estimates may be noisy. None of the datasets gave patient identifiers, so both eyes of one patient may fall into training and test parts. This can inflate the in-domain AUC, in particular for ACRIMA (0.998), and the reported drop may therefore be overestimated. Labeling rules differ between datasets and were not harmonized. Our own crops (2.5 times the disc size for ORIGA and G1020, automatic disc detection for some images) are mixed with the dataset differences. We did not test G1020 with uncropped images, so we cannot say whether its chance-level result comes from the labels, the crop or the pipeline.

*Statistics.* Three seeds give two degrees of freedom, so the confidence intervals are wide. The seeds share the same cross-source images, so the intervals describe training randomness and not variation between datasets. Pairwise model tests and normalization comparisons are exploratory, and the best normalization condition was identified only after seeing the results. We did not estimate uncertainty for the operating-point averages. The source threshold is poorly determined when the validation AUC is 1.0, as for most ACRIMA-source networks.

*Models and protocol.* We tested five widely used CNNs with fixed hyper-parameters and one 224 × 224 input size. Identical settings are not equal tuning. InceptionV3 was run at 224 × 224 rather than its native 299 × 299, and no architecture had its own hyper-parameter search. The low in-domain AUC of InceptionV3 (0.891) may partly reflect this, so differences between architectures should be read as specific to this protocol. Newer architectures were not tested, and the model comparison may depend on the protocol [42]. The recalibration experiment is a best case: labels are assumed available, calibration and evaluation images come from the same dataset, and only Youden's J was tried. The Grad-CAM maps are illustrative, from one model and eight images. At screening prevalence, predictive values would be much lower than our accuracy figures suggest.

## VI. Conclusion

We trained five ImageNet pre-trained CNNs on three public glaucoma fundus datasets and tested them in-domain and across datasets. The AUC fell from 0.913 to 0.702 (drop 0.211 ± 0.030; 95% CI 0.135 to 0.286). The thresholds chosen on the source data transferred poorly. Threshold recalibration on 10% of the target images raised specificity from 0.51 to 0.67, but it lowered sensitivity. The held-out AUC was not increased by just combining two sources. Changes of image appearance gave a partial recovery. LAB reference normalization and strong color augmentation gave similar gains in mean cross-source AUC (0.755, paired gain 0.053, 95% CI 0.046 to 0.060 across seeds; and 0.754), and Ben Graham normalization reached 0.733. LAB reference normalization reduced the drop by about 26%, but a large drop remained, and the higher specificity came with lower sensitivity at the source thresholds. CLAHE had a smaller effect, and gray-world normalization did not improve the average transfer. G1020 stayed near chance under every condition. Normalization and stronger augmentation are useful candidates for better cross-dataset performance. In-domain performance alone is not enough evidence that a model is ready for screening. Future work should test normalization together with multi-source training and target-specific operating points on independent centers. It should also include domain adaptation and careful control of the reference images.

## Code Availability

The training code, the data splits and the saved per-image scores are available in a public GitHub repository: https://github.com/alaminrahman33/glaucoma-cross-dataset-cnn.

## Acknowledgment

We thank Salman Fazle Rabby (Assistant Professor, Department of Electrical and Electronic Engineering, Sylhet Engineering College) for his supervision and guidance during this work.